# Empathy and the Human-Moment Gaps of AI Chatbots: Insights from Empathy Displacement Theory

**Victor Frimpong**
Management Department, SBS Swiss Business School, Kloten-Zurich, Switzerland.
v.frimpong@research.sbs.edu,
https://orcid.org/0009-0004-9560-3927

**Abstract:** *AI chatbots are increasingly deployed in domains where empathy is essential, such as healthcare, education, and customer service, yet their ability to sustain genuinely authentic "human moments" remains limited. This paper introduces two interlinked conceptual models to explain and address this limitation. The Human-Moment Gap Framework (HMGF) identifies three structural empathy deficits—affective surfaceism (emotional imitation without depth), memory fragmentation (lack of relational continuity), and moral framing mismatch (efficiency prioritised over dignity). Building on these mechanisms, the Empathy Displacement Theory (EDT) explains how AI-simulated empathy can progressively substitute, distort, and displace genuine human empathy at the individual, relational, and organisational levels. HMGF serves as the causal foundation of EDT, illustrating how technical and moral deficiencies in chatbot design evolve into broader social and institutional consequences. The paper is conceptual and exploratory, aiming to develop an integrative theoretical framework rather than to provide empirical validation. Together, the models provide a unified theoretical framework for understanding AI-mediated empathy, offering testable propositions and practical guidance for the ethically responsible development of chatbots. The paper closes with ethical and policy implications, arguing that the real challenge of empathetic AI lies not in whether machines can "care," but in how simulated care reshapes human affective expectations and institutional norms.*

## 1. Introduction

The exploration of empathy in conversational AI examines how technology shapes human feelings. Empathy is essential in healthcare, education, and customer service because it builds trust and nurtures strong connections (Pasupuleti, 2024; Shruti & Jyotsana, 2024; Lavrenchuk, 2025). Nevertheless, since AI chatbots mimic empathy rather than genuinely feel it, concerns arise about how this affects users' emotional expectations and social interactions.

Conversational AI offers efficiency and 24/7 accessibility, but doubts arise concerning the authenticity of its answers. Chatbots implement algorithms such as sentiment analysis and emotion recognition to simulate human empathy (Pasupuleti, 2024; Liu et al., 2025). While this imitation can boost user engagement, it may also alter the way users perceive empathy. Research indicates that users frequently attribute warmth and understanding to chatbots, potentially diminishing their interest in real human interaction (Alam et al., 2024). Moreover, even when AI delivers consistent service, its perceived empathy is shaped by how users interpret emotional signals (Feng et al., 2024).

Ethical issues surrounding AI are crucial. Relying on AI for emotional support can lead to dependence, which may weaken genuine human relationships (Lavrenchuk, 2025; Tolnaiová, 2024). Studies reveal that engaging with 'empathetic' AI can lessen prosocial behaviours and compassion towards others (Zhang et al., 2024). As AI takes on emotional and relational functions, it risks changing the very notion of empathy, transitioning from authentic moral engagement to mere algorithmic ease (Ansari, 2023; Ge, 2024).

To ensure AI enhances empathy rather than diminishes it, system design and governance must consider the complexities of human emotional communication. Integrating cognitive empathy into chatbot design can enhance their efficacy in initial triage or customer service, but it must be paired with safeguards to maintain genuine relationships and moral awareness (Alam et al., 2024; Basuki & Febrianta, 2024). Ultimately, to preserve authentic connections in a technology-driven world, it is crucial to focus on AI's functional potential while ensuring we do not lose the moral and emotional depth that defines humanity (Shruti & Jyotsana, 2024; Basuki & Febrianta, 2024).

This paper argues that AI systems cannot genuinely experience empathy because they lack consciousness, emotions, and moral awareness (Montemayor et al., 2021). However, they can simulate empathy through language and behaviour, which users perceive as emotionally responsive. Therefore, the analysis treats AI empathy as a simulated form that has real social consequences. The Human-Moment Gap Framework (HMGF) and Empathy Displacement Theory (EDT) focus on how this simulation, despite lacking genuine emotion, can reshape human expectations, moral perceptions, and institutional practices. This approach balances philosophical understanding with practical design by recognising that while machines cannot feel, their simulated empathy can still impact human relationships (Gabriel, 2020; Coeckelbergh, 2020).

### *1.1. Existing Approaches to AI Empathy*

Current research highlights the limitations and possibilities of artificial empathy. Philosophers argue that AI lacks the capacity for genuine empathy due to its limited consciousness and moral awareness (Montemayor, 2021). Design scholars emphasise the importance of creating frameworks to incorporate empathic cues into AI, facilitating more human-like interactions (Zhu & Luo, 2023). Studies have shown that in some situations, people may even find AI-generated responses more compassionate than those from humans (Ovsyannikova et al., 2025). However, managerial research reveals that while empathic chatbots can enhance customer satisfaction, they risk damaging trust if users detect inauthenticity (Juquelier et al., 2025).

These findings underscore three key points: AI's ability to empathise is limited, simulated empathy can be perceived as genuine, and this creates risks of overreliance and ethical misalignment. Despite these insights, existing research is fragmented. Many studies focus on isolated interactions rather than on ongoing relational dynamics. Others focus on specific aspects, such as accountability or design, without examining how AI influences the distribution of empathy at individual, relational, and organisational levels. Table 1 shows that no existing study integrates

AI's empathy deficits with a theory on how these issues might replace human empathy across different contexts.

### *1.2. The Missing Link: Human Moments and Displacement*

The Human-Moment Gap highlights a crucial oversight within the field: the difference between user expectations for empathy in interactions and what chatbots can actually provide. Users expect continuity, context, and moral sensitivity, but chatbots face limitations, including providing superficial emotional responses, fragmented memory, and prioritising efficiency over care. These limitations affect user trust and the dynamics of relationships, undermining the legitimacy of AI-driven services.

Moreover, when chatbots simulate empathy, they not only fail to meet expectations but can also alter them in unforeseen ways. This concept, known as Empathy Displacement, refers to how AI-simulated empathy can replace, distort, or displace genuine human empathy. Individually, users may rely on chatbots rather than seek human support (substitution). Relationally, exposure to chatbots may change expectations, making authentic human empathy seem inadequate (distortion). At the organisational level, companies might shift resources away from developing human empathy and lean more on AI for relational tasks (displacement).

We define this as the Empathy Displacement Theory (EDT), viewing it as an inevitable outcome of the Human-Moment Gap. Unlike previous studies that see AI empathy as either a design issue or a perception problem, EDT treats it as a fundamental shift in how empathy operates.

### *1.3. Contributions of this Paper*

This paper presents three key contributions—conceptual, theoretical, and practical—on how AI-simulated empathy influences human emotions and organisational dynamics.

- *Conceptual Contribution:* We introduce the Human-Moment Gap Framework (HMGF), which identifies gaps between users' expectations of empathy and what AI chatbots can provide. These gaps—affective surfaceism, memory fragmentation, and a mismatch in moral framing—explain why AI empathy can be shallow, inconsistent, and ethically contentious
- *Theoretical Contribution:* We propose the Empathy Displacement Theory (EDT), which details how simulated empathy can replace and distort genuine human empathy at individual, relational, and organisational levels. The HMGF generates structural empathy deficits that lead to this displacement.
- *Practical and Ethical Contribution:* Combining HMGF and EDT provides an integrated framework for research and practice. It includes (1) testable research propositions, (2) design principles for developing AI–human empathy systems, and (3) ethical guidance for organisations to foster genuine human connections amid the rise of emotional automation.

This study adopts a conceptual and theory-building approach because the phenomena under investigation—empathy simulation, relational expectation shifts, and institutional displacement—are emerging and not yet empirically established. Instead of testing predefined hypotheses, the goal is to develop integrative explanatory constructs that clarify underlying mechanisms and provide a foundation for future empirical research. Conceptual models like the HMGF and EDT function as theory-generating tools, offering structured frameworks through which subsequent studies can operationalise, measure, and test the dynamics of AI-mediated empathy.

## 2. Literature Review

This section reviews key scholarship on empathy as it relates to AI chatbots. We begin with definitions and types of empathy in psychology, then examine human–computer interaction (HCI) and affective computing. We also look at the role of empathy in professional and organisational settings. Finally, we identify gaps that led to the creation of the Human-Moment Gap Framework (HMGF) and Empathy Displacement Theory (EDT).

### *2.1 Empathy in Psychology: Foundations and Dimensions*

Empathy is a core psychological concept with three main components: affective, cognitive, and compassionate. Affective empathy entails emotionally connecting with others' feelings, cognitive empathy involves understanding their perspectives, and compassionate empathy drives the desire to help alleviate their suffering (Lișman & Holman, 2022). It plays a crucial role in doctor–patient relationships (Mercer & Reynolds, 2002), educational environments (Cooper, 2011), and conflict resolution (Batson, 2011). Empathy develops through interactions that foster trust and is assessed through words, timing, non-verbal cues, and shared experiences,rendering it difficult for technology to replicate.

The psychological literature identifies the components and interpersonal roles of empathy but assumes that both parties share emotional experiences. This assumption raises questions about how empathy operates when one party, such as an AI system, lacks emotional consciousness, which will be explored in later sections.

### *2.2. Empathy in Human–Computer Interaction and Affective Computing*

Research in HCI and affective computing aims to create systems that can recognise and respond to user emotions. Pioneering work, such as Picard's Affective Computing (1997), focused on giving machines emotion-recognition skills. This led to the development of chatbots and social robots which use sentiment analysis and natural language generation to convey empathy (Bickmore & Picard, 2005; Breazeal, 2003). Recent studies have shown that chatbots can be perceived as empathic; for instance, Ovsyannikova et al. (2025) found that some AI-generated responses were rated as more compassionate than human-generated responses. Additionally, patients may feel more comfortable sharing sensitive information with chatbots because they perceive them as less judgemental (Miner et al., 2016). However, researchers warn against "empathic illusions," where AI mimics empathy without genuine understanding (Montemayor, 2021). Two main limitations remain: most research focuses on short-term interactions, neglecting long-term trust, and evaluations mainly assess perceived empathy rather than its impact on human interactions and institutional dynamics over time

Most work in affective computing measures empathy based on perceived emotional appropriateness instead of true emotional understanding or ongoing relationships. This emphasis on short-term interaction overlooks how simulated empathy can shift expectations and norms of care over time, underscoring the need for the Human-Moment Gap Framework.

### *2.3. Organisational and Service Perspectives*

Empathy is recognised as essential in professional interactions beyond psychology and HCI. In healthcare, empathic communication leads to improved patient adherence, satisfaction, and health outcomes (Mercer & Reynolds, 2002). In customer service, it fosters loyalty and facilitates complaint resolution (Parasuraman et al., 1988). In education, teachers' empathy promotes student engagement and resilience (Cooper, 2011).

The implementation of chatbots brings both benefits and challenges. Juquelier et al. (2025) refer to empathic chatbots as a "double-edged sword" in customer service, as they can improve responsiveness but may fall short in terms of authenticity. Srinivasan (2022) raises concerns about accountability, questioning which standards ought to guide the use of AI that simulates empathy in sensitive areas. Zhu and Luo (2023) propose design frameworks for integrating artificial empathy into human-centred systems, with a focus on technical design rather than social implications. However, a significant gap exists regarding how organisations can replace human empathy with AI, which impacts organisational investments and professional roles.

Organisational studies usually focus on efficiency, consistent service, and trust repair, but they often overlook how empathy is redistributed between humans and machines. This lack of focus on emotional labour highlights the necessity for a theory that connects individual interactions to broader displacement effects at meso and macro levels.

### 2.4. Identified Gaps

Synthesising across these literatures, three significant gaps become apparent.

1. ***Relational continuity:*** Most studies concentrate on single interactions, but there is limited research on how perceptions of empathy change in repeated encounters with AI, particularly when continuity of memory is disrupted.
2. ***Displacement effects:*** There is no formal theory explaining how AI's simulated empathy can replace, distort, or displace genuine human empathy, despite recognising the lack of empathy in AI.
3. ***Multi-level integration:*** Current studies separate empathy into psychological (individual) and design (technical) areas, with few that integrate user-level, relational, and organisational aspects into a comprehensive framework.

We propose the Human-Moment Gap Framework (HMGF) to identify barriers between user expectations and chatbot capabilities. Additionally, we present the Empathy Displacement Theory (EDT), which explores the multi-level consequences of AI-simulated empathy.

To highlight the novelty, Table 1 summarises related work and contrasts it with the contributions of this paper.

*Table 1. Prior work on AI empathy vs. this paper's contribution*

| Author(s) & Year | Focus / Contribution | Key Limitation / Gap | Relevance to Present Study |
|---|---|---|---|
| Montemayor (2021) | Philosophical barriers to artificial empathy | Discusses limits but not relational continuity or displacement effects | Supports HMGF's "Affective Surfaceism" |
| Srinivasan (2022) | Empathy and AI accountability | Focuses on accountability; no displacement theory | Highlights ethical urgency but not structural dynamics |
| Juquelier et al. (2025). | Empathic chatbots in customer service | Managerial trade-offs; lacks longitudinal theory | Shows practical risks but not theorised displacement |
| Zhu & Luo (2023). | Framework for artificial empathy in design | Engineering-oriented; no account of expectation distortion | Background for design, not displacement |
| Ovsyannikova et al. (2025). | AI is sometimes rated as more compassionate than humans. | Short-term perception; no longitudinal analysis | Supports EDT's substitution mechanism |
| Mahnaz Roshanaei et al., 2024 | ChatGPT vs. human empathy in tasks | Static comparisons; ignore long-term trust | Supports "Memory Fragmentation" gap |
| **This paper (2025)** | Introduces HMGF + EDT | — | The first to integrate gaps and multi-level displacement. |

Empathy is well-explored in psychology and applied in HCI and service design, but no current framework connects human-moment gaps to a multi-level displacement theory. This paper addresses that gap by presenting the HMGF and EDT as novel theoretical contributions for advancing empirical, conceptual, and policy research on AI and empathy.

### 2.5. Methodological Approach

This paper adopts a conceptual and integrative approach to AI-mediated empathy, focusing on theoretical development for clarity and replicability. Its goal is to unify scattered research from psychology, human–computer interaction, and organisational studies into a clear framework.

The review and development of the model were conducted in three stages:

1. A targeted theoretical sampling approach (Webster & Watson, 2002) was employed to identify recent, high-impact studies (2016–2025) on AI, empathy, emotional labour, and moral alignment. Searches were performed in Scopus, Web of Science, and Google Scholar using search terms such as "AI empathy," "chatbots," "affective computing," "moral alignment," and "emotional simulation."

2. A conceptual synthesis approach (Jaakkola, 2020) was applied to integrate insights from various disciplines. This led to the identification of three key empathy deficits—affective surfaceism, memory fragmentation, and moral framing mismatch—which collectively constitute the Human-Moment Gap Framework (HMGF).
3. Theoretical integration connected these gaps to broader systemic outcomes through cross-level reasoning, leading to the development of the Empathy Displacement Theory (EDT).

## 3. The Human-Moment Gap Framework (HMGF)

AI chatbots are now widely used in critical areas such as healthcare triage, mental health support, classroom tutoring, and customer service, where empathy is essential. However, users often find AI empathy deficient compared to human interactions. To explain this empathic disconnect, we present the HMGF, which identifies three main gaps: affective surfaceism, memory fragmentation, and moral framing mismatch. These gaps reveal the limitations of technology and highlight the differences between human expectations of care and what machines can provide.

### *3.1. Affective Surfaceism*

Affective surfaceism occurs when AI systems mimic emotional responses superficially—using empathic language, emojis, or tone markers —but lack an accurate understanding or emotional depth. While users may initially perceive these responses as empathic, they quickly realise that the AI does not genuinely understand their feelings.

In mental health chatbots, generic responses like “I am sorry to hear that. That must be hard for you,” lack context. They do not address if a user is experiencing grief, anxiety, or anger. Over time, these expressions of empathy can appear inauthentic, a phenomenon known as empathic illusions, as described by Montemayor (2021). While such responses may provide temporary comfort, they ultimately widen the gap between human and machine empathy.

### *3.2. Memory Fragmentation*

Memory fragmentation occurs when chatbots fail to maintain continuity between interactions. Unlike humans, who remember personal details to express care, chatbots usually retain contextual information only within a single session and forget past details once the conversation ends.

In online education, a tutoring chatbot can provide pedagogically relevant explanations in a single session but often fails to recall a student’s past misconceptions in subsequent interactions. This inconsistency undermines the bond that human teachers build through familiarity. Fragmented memory disrupts the trust needed for empathy. Even advanced systems with persistent memory face challenges related to consent, privacy, and selective recall, which complicate the delivery of consistent and empathic support.

### *3.3. Moral Framing Mismatch*

Moral Framing Mismatch occurs when AI systems prioritise efficiency and compliance over human empathy and moral values. This misalignment highlights a significant issue in AI ethics, where decision-making does not reflect the ethical standards humans expect in sensitive situations. According to Moral Disengagement Theory (Bandura, 1999), this parallels how individuals justify harmful actions by viewing them as necessary or neutral.

For instance, a healthcare chatbot that prioritises patients based on statistical urgency over emotional distress reduces people to mere data points. This undermines empathy and leads to responses that may be effective but are devoid of moral depth. Users can perceive when systems prioritise optimisation over dignity and care, eroding the essence of empathy.

Recent research highlights the importance of aligning moral values in AI ethics. Gabriel (2020) contends that alignment frameworks need to go beyond performance measures to encompass the ethical consequences of decisions that affect humans. When empathy is treated merely as a metric for efficiency, it can lead to moral insensitivity, perceiving relational care as an expense rather than a moral obligation. In the end, this disconnect jeopardises trust and credibility in areas where empathy is vital, such as healthcare, education, and social services.

### *3.4. Integrative Perspective*

The three gaps underscore the challenges AI chatbots face in fostering genuine human interaction. Affective surfaceism limits empathy to superficial imitation, while memory fragmentation hinders continuity and the development of meaningful relationships. Additionally, a disparity in moral framing prioritises operational efficiency over dignity.

The framework (Figure 1) depicts the factors that contribute to a persistent empathy deficit between user expectations and chatbots' capabilities. Significantly, these issues extend beyond mere technical constraints; they also lead users to adjust their expectations, behaviours, and organisational practices in response to AI that only partially satisfies their empathic needs. The following section will introduce the EDT to explore further how these changes affect broader structural transformations in the context of empathy.

This illustration emphasises three deficiencies in empathy that emerge when human emotional expectations clash with the limitations of AI chatbots: affective surfaceism (superficial emotional replication), memory fragmentation (lack of relational continuity), and moral framing mismatch (favouring efficiency over dignity). These deficiencies contribute to the Human-Moment Gap, the disconnect between genuine human empathy and the emotional simulations AI offers. This framework demonstrates how the constrained empathic responses from chatbots can shift emotional expectations, setting the stage for the subsequent displacement effects depicted in Figure 2.

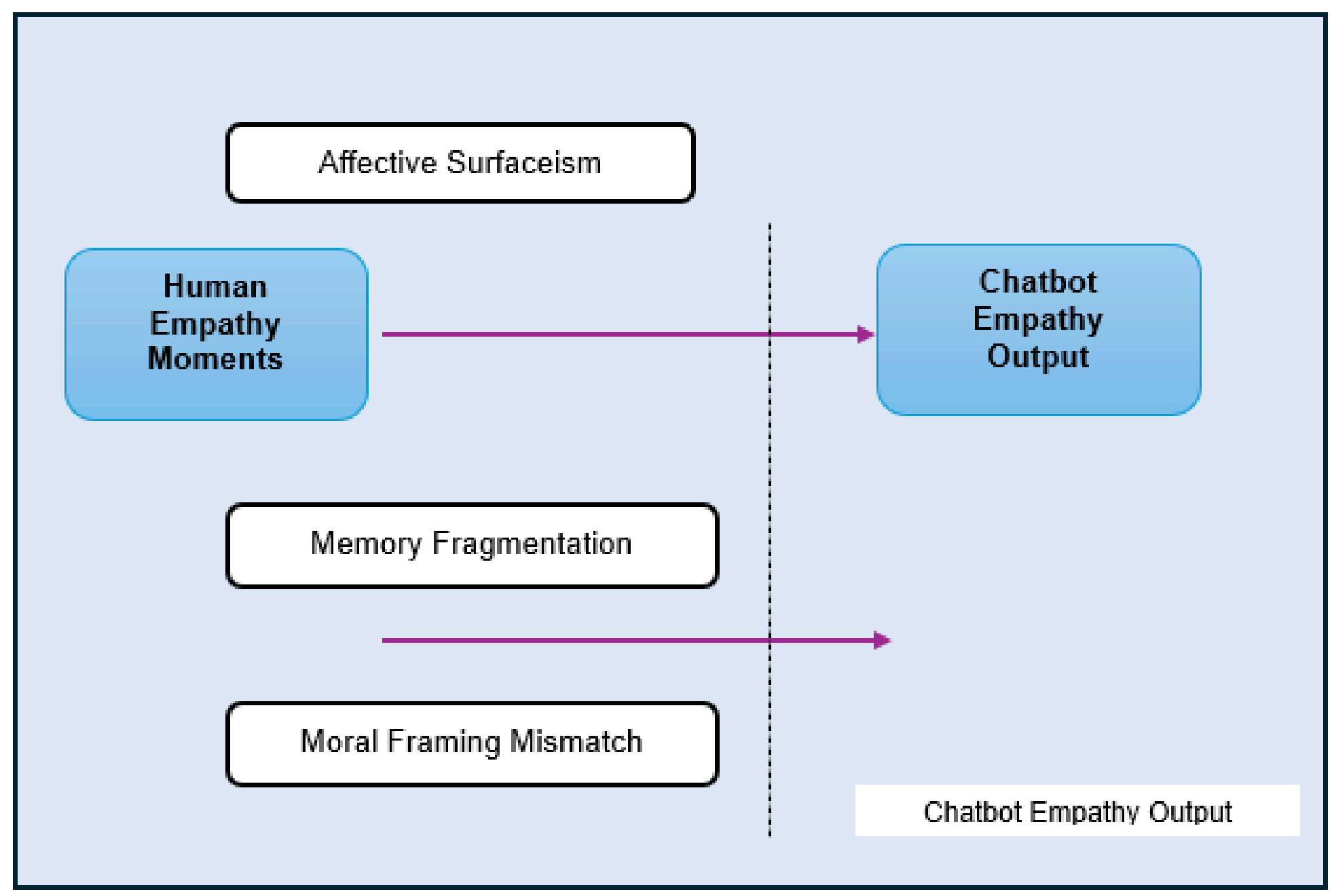

*Figure 1. Human-Moment Gap Framework (HMGF)*
Source: the author, 2025

The Human-Moment Gap highlights a lack of AI empathy and explains how this leads to a shift in genuine empathy. Three specific gaps are driving users, relationships, and organisations to rely more heavily on artificial empathy. Affective surfaceism leads individuals to seek predictable, but shallow, comfort from chatbots.

Memory fragmentation may result in distorted relationships, altering how people perceive and value empathy. Moral framing mismatches lead organisations to prioritise efficiency over genuine care. Together, these issues create an empathy deficit that evolves into a displacement effect, in which imitation gradually alters how empathy is perceived and exchanged. This is outlined in the Empathy Displacement Theory (EDT), which illustrates how simulated empathy moves from being a gap to a replacement through ongoing feedback loops.

*Table 2. Linking HMGF gaps to EDT levels: the causal bridge*

| Human-Moment Gap (HMGF) | Nature of the Gap | Resulting Displacement Level (EDT) | Mechanism of Transition | Illustrative Example |
|---|---|---|---|---|
| **Affective Surfaceism** | AI mimics emotional language but lacks genuine understanding or affective depth. | **Individual Substitution** | Users increasingly prefer chatbots' predictable emotional responses and simulated empathy over human connection. | A user repeatedly consults a wellness chatbot for comfort instead of reaching out to a friend or counselor. |
| **Memory Fragmentation** | AI struggles to maintain relational continuity, often forgetting prior context or disclosures. | **Relational Distortion** | Inconsistent memory alters empathy expectations, making genuine human empathy seem messy or inefficient. | A student using a tutoring bot finds human teachers to be "inconsistent" after having perfect chatbot sessions. |
| **Moral Framing Mismatch** | AI prioritises efficiency, compliance, or risk reduction over dignity and moral care. | **Organisational Displacement** | Institutions shift empathy work from humans to machines, embedding moral detachment into systems. | A healthcare provider replaces initial patient interviews with AI triage bots, reducing the need for human dialogue. |

The mapping presented in Table 2 shows that Human-Moment Gaps are not static constraints but active variables that induce shifts in empathy. Each gap triggers a corresponding structural adjustment—whether at the individual, relational, or organisational level—that reorients empathy from humans toward AI systems. This causal mechanism elucidates how interactional gaps (HMGF) evolve into systemic displacement (EDT), thus linking micro-level communication issues with larger social transformations.

## 4. Empathy Displacement Theory (EDT)

The HMGF points out the constraints of AI chatbots in delivering genuine human empathy; nevertheless, it overlooks the broader implications of simulated empathy. To address this oversight, we introduce the EDT, which describes how AI-facilitated empathy can substitute for, alter, and eventually supplant genuine human empathy.

EDT is based on three intellectual traditions:

1. Social Presence Theory (Short, Williams & Christie, 1976) explores how various media shape perceptions of authenticity in communication.

2. The Displacement Hypothesis in media research (Nie, 2001) suggests that the use of new media diverts time that would otherwise be spent on traditional face-to-face interactions.

3. Moral salience theories in ethics (Darwall, 2006) highlight that empathy comes with normative responsibilities — a capability that machines cannot wholly replicate.

By bringing these components together, EDT argues that AI empathy is not just absent but also has the power to transform environments: it changes how individuals experience empathy, their relationship expectations, and how organisations allocate it as a resource.

The transition from the Human-Moment Gap Framework (HMGF) to the Empathy Displacement Theory (EDT) represents a process-oriented causal mechanism rather than a simple descriptive relationship. The structural empathy deficits identified in HMGF—such as affective surfaceism, memory fragmentation, and moral framing mismatch—lead to predictable patterns of user adaptation. When users engage consistently with simulated empathy that is emotionally coherent yet affectively shallow, they tend to recalibrate their expectations regarding the experience of empathy. This recalibration initiates three sequential transformations: individuals increasingly replace human support with chatbot interactions, relational norms shift to prioritise efficiency and emotional predictability, and organisations subsequently reallocate empathy-related tasks from human to machine. EDT conceptualizes empathy displacement not as a singular event but as a cumulative feedback process driven by ongoing exposure to simulated empathy amid unfulfilled human emotional expectations.

### *4.1. Levels of Displacement*

***Individual-level substitution:*** Individuals may increasingly seek emotional comfort and cognitive guidance from chatbots rather than human empathy. Due to their always-on availability, lack of judgement, and reliability, chatbots might appear to be "safer" alternatives. This dependency could result in fewer personal interactions, diminishing reliance on family, friends, or professionals. For example, young people might prefer to discuss their problems with a mental health chatbot rather than a teacher or parent, because the chatbot is more approachable.

***Relational-Level Distortion:*** At the meso level, frequent interactions with empathic chatbots can alter users' expectations of what empathy entails. If chatbot responses are always immediate, consistent, and refined, users may perceive human empathy as messy and inadequate. This can negatively affect their tolerance for human flaws and diminish their appreciation for genuine, yet imperfect, empathic responses. For instance, customers accustomed to quick chatbot replies, such as "I understand your frustration, let me fix this," may feel disappointed by a human agent who responds more awkwardly or takes longer to express empathy.

***Organisational-Level Displacement:*** Organisations may shift resources based on how well chatbots demonstrate empathy. If chatbots effectively satisfy customers or patients, companies may reduce their spending on training employees in empathic communication or eliminate roles focused on emotional labour. This can lead to a decline in genuine human empathy within institutions. For example, a hospital using chatbots for initial patient intake may lower the number of staff responsible for direct patient communication, changing the overall care environment.

### *4.2. Integrative Model*

The EDT, as illustrated in Figure 2, describes a process in which AI-simulated empathy can lead to individual replacement, relationship distortion, and organisational shift. This model shows how simulated empathy turns into structural change across three levels of human experience. At the individual level, people substitute human interaction with chatbot engagement (replacement). At the relationship level, frequent exposure to polished but emotionless empathy changes expectations of genuine care (distortion). At the organisational level, organisations increasingly assign empathic tasks to AI systems (shift). Feedback loops reveal that these levels support and reinforce one another, gradually making artificial empathy an everyday organisational practice. Factors such as domain risk, cultural norms, and system design influence the intensity and reversibility of this shift.

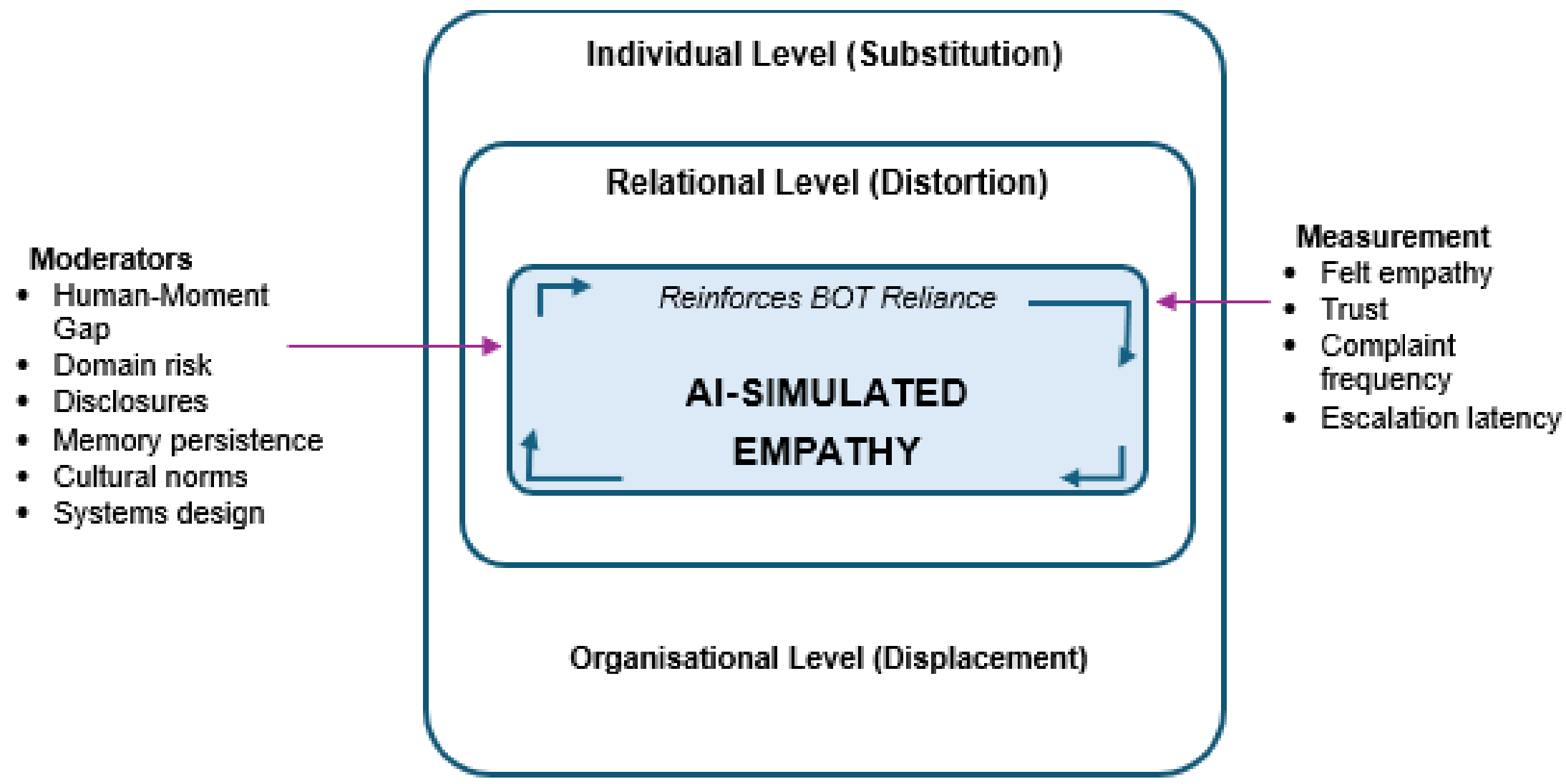

*Figure 2. Empathy Displacement Theory (EDT): multi-level model*
Source: the author, 2025

### *4.3 Propositions for Research*

To make EDT empirically testable, we propose the following observable propositions:

- ***P1 (Substitution):*** Higher perceived empathy in chatbots leads to a lower probability of individuals seeking empathic support from humans in the same area.
- ***P2 (Distortion):*** Frequent interactions with empathic chatbots can lower people's expectations for genuine human empathy, leading them to perceive even genuine but flawed human empathy as less satisfying over time.
- ***P3 (Displacement*):** Organisations that deploy empathetic chatbots extensively will face reduced expenses for human empathy services in comparison to those that do not.
    - Research direction: The phenomenon of organisational-level displacement can be examined through comparative case studies across various sectors. For example, we could analyse healthcare organisations or customer service companies that implement empathetic chatbots versus those that rely exclusively on human personnel. Important metrics to evaluate include budgets for communication skills training, staff-to-client ratios, and changing descriptions of employee roles. Surveys directed at managers can assess their perceptions of using AI for emotional labour. At the same time, archival information, such as staffing trends and financial data, can provide objective measures of resource reallocation..
- ***P4 (Gap–Displacement Link):*** The extent of the Human-Moment Gap influences the degree to which empathy is replaced; greater gaps result in increased substitution, distortion, and displacement within organisations.

  o Research direction: Scholars can investigate the Human-Moment Gap by exploring its three primary dimensions: affective surfaceism, memory fragmentation, and moral framing mismatch. This investigation can utilise established scales and coding techniques. For example, transcripts of conversations can be evaluated for empathic depth, continuity indicators, and moral framing cues. Correlational analyses can then identify whether larger perceived gaps are linked to greater displacement effects, such as increased dependence on chatbots, diminished expectations for human interaction, or reductions in organisational resources. An integrated approach could merge survey data with ethnographic studies within service or clinical environments.
- ***P5 (Feedback Loop):*** Organisational dislocation results in individual substitution and skewed relationships, forming a cycle that diminishes empathy in environments shaped by AI.

EDT develops the existing theory in three distinct ways. First, it draws on the displacement hypothesis from media studies to examine AI empathy, emphasising the dangers of diminishing moral and relational skills. Second, it elaborates on social presence theory by demonstrating the ways in which "artificial presence" can not only imitate but also change relational standards. Third, it presents an ethical viewpoint, cautioning that prioritising organisational efficiency might undermine authentic human-centred care.

Table 3 maps the main elements linking the Human-Moment Gap Framework (HMGF) and the Empathy Displacement Theory (EDT). It defines key terms, identifies analytical levels, explains mechanisms, and provides applied examples for better understanding and practical use.

*Table 3. Summary of Key Constructs and Theoretical Linkages*

| Construct | Definition | Level of Analysis | Related Mechanism | Illustrative Example |
|---|---|---|---|---|
| **Affective Surfaceism** | Superficial imitation of emotional cues like tone, emojis, or comforting phrases, lacking genuine understanding or depth. | *Individual Level* | **Substitution –** Users tend to rely more on predictable simulated empathy than on real human connection. | A wellness chatbot repeatedly says, "I'm sorry to hear that," encouraging users to confide in it instead of a friend or counselor. |
| **Memory Fragmentation** | AI systems cannot maintain relational continuity or recall past interactions, disrupting trust and familiarity. | *Relational Level* | **Distortion—**fragmented memory alters expectations, making genuine empathy seem inconsistent or inefficient. | A tutoring chatbot forgets a student's earlier struggles, yet the student starts to see human teachers as disorganised by comparison. |
| **Moral Framing Mismatch** | Prioritising efficiency, compliance, or risk reduction over moral sensitivity and dignity in AI decision-making. | *Organisational Level* | **Displacement –** Institutions shift empathy work to machines, embedding moral detachment into routines. | A hospital triage bot assigns urgency based on statistics rather than distress, which decreases moral engagement in providing care. |
| **Empathy Displacement Theory (EDT)** | The process by which simulated empathy gradually replaces, distorts, and displaces genuine human empathy at various levels. | *Cross-Level / Systemic* | **Feedback Loop –** Reliance fosters dependence, normalising artificial empathy. | Routine chatbot interactions make users and organisations see simulated empathy as normal. |

## 5. Propositions and Research Directions

The EDT outlines how AI-generated empathy affects human interactions at the individual, relational, and organisational levels. It presents key propositions and methods for empirical testing.

### Individual substitution

- ***Proposition 1 (Substitution):*** An increased perception of empathy in chatbots reduces the chances of people looking for empathic assistance from humans in the same context.
  - Research direction: This hypothesis can be empirically tested through experiments and surveys. Participants will interact with a neutral chatbot, an "empathic" chatbot, or a human. We will compare their willingness to seek help through contacting a counsellor, doctor, or friend across these groups. Additionally, long-term studies in mental health apps or educational tutoring will evaluate whether regular chatbot use reduces reliance on human teachers, therapists, or mentors.

**Relational distortion**

- ***Proposition 2 (Distortion):*** Frequent interactions with empathic chatbots can lower people's expectations for genuine human empathy, causing them to perceive even genuine but flawed human empathy as less satisfying over time.
  - Research direction: This proposition supports longitudinal studies that track users across multiple sessions to measure their satisfaction with both chatbots and human interactions. Participants can be exposed to different levels of chatbot empathy (high vs. low in consistency, immediacy, and linguistic polish) to assess their evaluations of human responders in later interactions. Qualitative interviews can further complement these findings by investigating how participants describe their changing perceptions of empathy.

**Organisational displacement**

- ***Proposition 3 (Displacement):*** Organisations using empathic chatbots at scale will spend less on human empathy services than those that do not.
  - Research direction: Organisational-level displacement can be analysed through comparative case studies across different industries. For instance, we can compare healthcare institutions or customer service providers that use empathic chatbots with those that depend solely on human staff. Key metrics for consideration include training budgets for communication skills, staff-to-client ratios, and evolving employee role descriptions. Manager surveys may assess attitudes toward using AI for emotional labour, while archival data such as staffing patterns and financial reports can reveal objective indicators of resource reallocation.

**Gap–displacement link**

- ***Proposition 4 (Gap–Displacement Link):*** The size of the Human-Moment Gap determines how much empathy is displaced; larger gaps lead to more substitution, distortion, and organisational displacement.
  - Research direction: Researchers can study the Human-Moment Gap by examining its three main dimensions: affective surfaceism, memory fragmentation, and moral framing mismatch. This can be done using established scales and coding methods. For instance, conversation transcripts can be analysed for empathic coding, continuity markers, and moral framing signals. Correlational analyses can then determine whether larger perceived gaps are associated with higher displacement outcomes, such as increased reliance on chatbots, lower expectations for human interaction, or cuts in organisational resources. A mixed-methods approach could combine survey data with ethnographic observations in service or clinical settings.

**Feedback loop dynamics**

- ***Proposition 5 (Feedback Loop):*** Organisational displacement leads to individual replacement and distorted relationships, creating a cycle that reduces empathy in AI-driven environments.
  - Research direction: This proposal suggests using multilevel modelling to analyse longitudinal organisational data alongside individual surveys. By examining how staff reductions in empathic roles impact user reliance on chatbots, we can justify further organisational downsizing. Additionally, simulation modelling or system dynamics can highlight the feedback loops between individual behaviours and broader organisational trends.

The feedback loop created by replacing human empathy with AI systems is self-perpetuating. As users increasingly engage with chatbots, this legitimises the ongoing replacement of human empathy, leading to a rapid decline in genuine empathy. This creates a path dependency on simulated empathy as the norm. However, this trend is not irreversible. With the right regulatory, cultural, or design changes—such as mandatory human-empathy quotas in care settings or clear disclosure protocols—it is possible to restore balance and highlight the value of genuine human empathy.

### *5.1. Measurement and Conceptual Testability*

The propositions in this study are theoretical, but they can be operationalised through empirically observable indicators. Table 4 presents illustrative examples of potential measures for future empirical research to validate the Empathy Displacement Theory (EDT). These examples serve as starting points for turning abstract concepts into measurable variables.

*Table 4. Illustrative indicators for testing the Empathy Displacement Theory*

| Construct | Possible Measurable Indicators | Measurement Context / Method |
|---|---|---|
| *Affective Surfaceism* | Frequency of generic empathetic phrases ("I understand," "That must be hard"); sentiment-analysis scores showing emotional uniformity | Text-mining of chatbot conversation logs; linguistic coding |
| *Memory Fragmentation* | Ratio of continuity references (e.g., recalling prior user details) to total interactions; user-reported sense of relational continuity | Session-tracking data; longitudinal user surveys |
| *Moral Framing Mismatch* | Degree to which AI prioritises efficiency metrics over moral cues in decision logs; user perception of dignity or fairness | Audit of AI decision trees; experimental vignette surveys |
| *Empathy Substitution (Individual Level)* | Reduction in frequency of seeking human support after repeated AI interaction | Pre-/post-interaction behavioural logs; self-report diaries |
| *Empathy Distortion (Relational Level)* | Decline in satisfaction with human empathy following frequent chatbot use | Panel survey of users over time |
| *Empathy Displacement (Organisational Level)* | Changes in training budgets or staffing for empathy-related roles after chatbot adoption | Archival HR and financial data; managerial interviews |

Table 4 outlines how to measure the key elements of HMGF and EDT, showing how each proposition (P1–P5) can be evaluated using specific indicators at individual, relational, and organisational levels.

### *5.2. Methodological Pluralism*

Testing the Empathy Displacement Theory requires a multilevel approach to adequately capture its complex, dynamic nature effectively. Since empathy displacement occurs at individual, relational, and organisational levels, no single method can fully address it.

Controlled experiments can quantify causal mechanisms, such as substitution and distortion, to assess how simulated empathy affects human responses. Longitudinal field studies can further explore how repeated interactions with empathic chatbots influence expectations of human empathy over time. Organisational case studies can highlight structural and cultural changes due to the widespread use of AI-mediated empathy.

Combining these methods—experimental, longitudinal, and organisational—provides a comprehensive understanding of how empathy is experienced, adapted, and institutionalised across different contexts. This methodological pluralism enhances the empirical testing of EDT while accounting for the moral, relational, and systemic dimensions of AI-mediated empathy.

A mixed-method and multi-level research strategy is essential for studying empathy displacement. The constructs in Empathy Displacement Theory (EDT) operate on different levels: individual cognition, relational interactions, and organisational resources. These levels cannot be effectively captured with just one data source or method. Quantitative methods, like experiments and behavioural metrics, are needed to identify causal effects and measure changes in empathy perception. Meanwhile, qualitative methods, such as interviews and case studies, help explain how empathy norms are interpreted and adopted over time. Using a mixed-methods approach allows researchers to explore both the underlying mechanisms and the real-life experiences of empathy displacement, matching the research method to the theory's multi-level nature.

### *5.3. Effects on Empirical Validation*

Testing EDT will validate its claims and guide design, ethics, and policy. For instance, if substitution effects are significant, mental health and education systems should create hybrid interventions that prompt users to seek human support. If distortion effects occur, user education and clear disclosure protocols may be needed to manage expectations about human empathy. If organisational displacement is noted, regulators may need to require a minimum level of human empathy in critical areas, such as healthcare and elder care.

### *5.4. Summary of Hypotheses*

Figure 3 outlines the Empathy Displacement Theory (EDT). It shows how Human-Moment Gap factors—Affective Surfaceism, Memory Fragmentation, and Moral Framing Mismatch—impact Simulated Empathy Output.

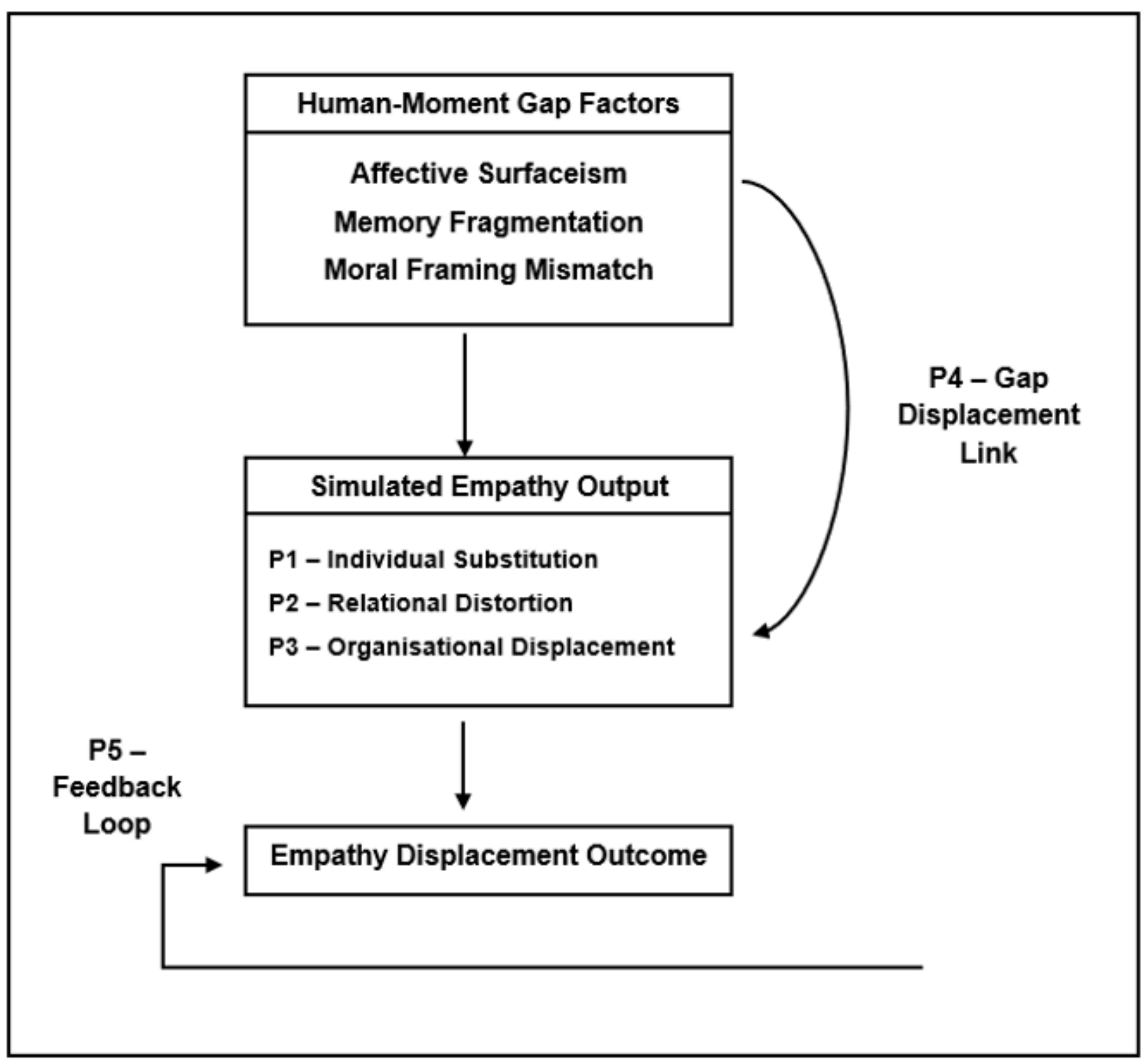

*Figure 3. Schematic overview of the Empathy Displacement Theory (EDT) propositions*
Source: the author, 2025

This output leads to three displacement mechanisms: Individual Substitution (P1), Relational Distortion (P2), and Organisational Displacement (P3). The strength of these relationships is influenced by the size of the Human-Moment Gap (P4), and feedback dynamics (P5) continue to shape human expectations and reliance on artificial empathy.

## 6. Implications

HMGF and EDT offer valuable insights for developing theory, informing managerial decision-making, and designing AI-mediated empathy-related policies.

### *6.1. Theoretical Implications*

This framework views AI empathy as a functional simulation rather than emotional equivalence. The ethical and organisational impacts come not from any feelings AI may have but from how effectively it simulates empathy and how humans recalibrate their emotional standards in reaction to it.

***Extending Empathy Scholarship***

HMGF enhances empathy research by focusing on structural mismatches instead of just deficits. It introduces terms such as affective surfaceism, memory fragmentation, and moral framing mismatch, providing a vocabulary for scholars to identify empathy gaps in AI-mediated interactions.

***Advancing Displacement Theory***

EDT expands on the media displacement hypothesis by examining how technology affects not only time use but also our relational and moral capacities. It argues that empathy is not a fixed trait, but a social resource that can be reshaped or diminished by technology.

***Integrating multi-level analysis***

EDT provides a unique cross-scale theory by addressing individual (substitution), relational (distortion), and organisational (displacement) levels. It integrates micro-level psychology, meso-level relationship dynamics, and macro-level organisational structures. Future research may employ this multilevel approach to explore other forms of "empathy work" across areas such as politics, education, and digital communities.

### *6.2. Managerial Implications*

***Designing hybrid empathy systems***

Managers ought not to regard chatbots as replacements for human empathy. Instead, they should adopt hybrid models in which chatbots handle routine queries and escalate emotional matters to human staff. For instance, a banking chatbot can handle transactions but should direct calls about death and grief to trained employees.

***Investing in continuity and memory***

To tackle memory fragmentation, organisations should implement systems that preserve user context across sessions and clearly communicate memory usage, allowing users to opt out if desired. Even recalling recent issues can enhance empathy and minimise relational breakdowns.

***Training employees for complementary empathy***

If reduced staffing due to displacement pressures leads to fewer empathic employees, those who remain will bear greater emotional burdens. Training should focus on how humans can provide empathy in ways AI cannot, such as nuanced judgement and moral recognition. This may redefine human roles into "escalation specialists" or "empathy anchors."

***Monitoring unintended consequences***

Organisations implementing empathic chatbots must monitor for substitution and distortion effects. Regular user feedback, complaint analysis, and trust audits can help detect if AI empathy is overshadowing human interactions, allowing managers to resolve issues promptly.

### *6.3. Policy and Ethical Implications*

***Disclosure and transparency***

Policymakers ought to require that systems acknowledge their limits in empathy. For example: "I am an AI assistant. While I can offer empathic replies, I might not completely understand your feelings. If you require human assistance, click here." This clarification helps prevent misconceptions about the system's abilities.

***Safeguards in high-stakes contexts***

In healthcare, mental health, and elder care, regulations typically mandate a minimum standard of human interaction. Just as aviation law requires human pilots even with automation, industries that involve moral considerations often need to ensure human empathy.

***Accountability for displacement.***

If organisations cut back on empathic staffing because of AI, regulators and unions might demand audits to ensure care quality remains high. Policies should require documenting investments in both AI and human empathy to maintain service standards.

***Cross-cultural considerations***

Policies must recognise that cultural differences affect expectations of empathy and compassion. Collectivist societies prioritise relationship continuity, whereas individualist societies value quick responses. Policymakers should account for these socio-cultural variations when implementing AI-driven empathy.

***Mini-Case Examples of Ethical and Policy Issues***

- *Healthcare Case – AI Triage and Empathy Substitution.*
  A large European hospital implemented an AI triage chatbot to prioritise emergency patients. While this improved waiting-room efficiency by a significant percentage, patient satisfaction decreased because users felt "unheard." This case shows that focusing on efficiency can undermine relational empathy. It suggests the need for policies requiring a human-in-the-loop approach, ensuring that at least one empathic interaction involves a clinician.
- *Education Case – Learning Companion Bots.*
  In a pilot at an online university, a tutoring bot offered continuous support with phrases like "I believe in you!" Students felt more motivated but became less dependent on their peers. This indicates a distortion in relationships, leading to the recommendation that regulators mandate reminders that the bot is simulating empathy.
- *Customer-Service Case – Emotional Labour Automation.*
  A retail bank replaced its frontline staff with chatbots that use sentiment analysis. While complaints decreased in the short term, trust scores declined when customers learned about the automation. This highlights the importance of transparency in AI and the need for audit frameworks to monitor the outsourcing of empathy.

HMGF and EDT have important implications. The mini-cases illustrate how these concepts can be applied in authentic contexts, guiding ethical design and regulatory decisions. They deepen our understanding of empathy and displacement, provide strategies for managing hybrid systems, and stress the importance of safeguards to protect empathy. Ultimately, AI empathy is not just a design concern; it signifies a fundamental shift in how societies distribute, experience, and value empathy.

## 7. Conclusion

Empathy is generally regarded as a fundamental human trait essential for fostering trust and care in relationships. However, as AI chatbots become increasingly common across sectors such as healthcare, education, and customer service, the ability of these systems to simulate empathic behaviour has emerged as a significant concern. While previous studies have explored empathy in psychology and organisational contexts, they often overlook the broader implications of simulated empathy.

This paper introduces two key concepts: the Human-Moment Gap Framework (HMGF) and the Empathy Displacement Theory (EDT). HMGF identifies barriers that hinder chatbots from creating genuine human connections, including affective surfaceism, memory fragmentation, and a mismatch in moral framing. EDT explores the consequences of these limitations, revealing how AI-simulated empathy can replace human interactions at the individual level, skew relational expectations, and shift the burden of empathy within organisations.

These contributions deepen our understanding of empathy and its displacement, yielding testable hypotheses for research and offering critical insights for managers and policymakers. The core idea is that the issue with AI empathy extends beyond whether machines can "care" to encompass how this simulated care affects human empathy dynamics. By defining the Human-Moment Gap and Empathy Displacement, this paper provides valuable tools for anticipating and addressing the loss of authentic human interactions in a world increasingly mediated by AI.

While the paper highlights the risks of empathy displacement, it is important to recognise that AI-mediated empathy can complement human empathy in certain situations. In areas such as mental health, education, and customer care, chatbots can provide first-contact support, increasing access to help and easing the burden on professionals. When designed transparently and positioned as assistive tools, AI empathy can enhance human connection by encouraging users to seek further help and reinforcing empathic behaviours among staff.

Additionally, the trends of displacement are not permanent. The same design and governance approaches that may diminish empathy can also restore it. Implementing human-in-the-loop protocols, providing empathy training for AI supervisors, and mandating the disclosure of emotional simulation can create systems in which machines support rather than replace human moral engagement. Ultimately, the future of AI-mediated empathy depends not just on technology but also on the ethical intentions and institutional frameworks that guide its use.